\documentclass{Interspeech2024}
\usepackage{multirow}
\usepackage{subfigure}

\interspeechcameraready

\title{Exploring synthetic data for cross-speaker style transfer in style representation based TTS}

\name[affiliation={1,2}]{Lucas H.}{Ueda}
\name[affiliation={1,2}]{Leonardo B. de M. M.}{Marques}
\name[affiliation={2}]{Flávio O.}{Simões}
\name[affiliation={2}]{Mário U.}{Neto}
\name[affiliation={2}]{Fernando}{Runstein}
\name[affiliation={2}]{Bianca}{Dal Bó}
\name[affiliation={1}]{Paula D. P.}{Costa}

\address{
  $^1$ Universidade Estadual de Campinas (UNICAMP), $^2$ CPQD, Brazil}
\email{paulad@unicamp.br,lhueda@cpqd.com.br}

\keywords{text-to-speech, cross-speaker style transfer, expressive speech synthesis, synthetic data, representation learning}

\begin{document}

\maketitle

\begin{abstract}

Incorporating cross-speaker style transfer in text-to-speech (TTS) models is challenging due to the need to disentangle speaker and style information in audio. In low-resource expressive data scenarios, voice conversion (VC) can generate expressive speech for target speakers, which can then be used to train the TTS model. However, the quality and style transfer ability of the VC model are crucial for the overall TTS model quality. In this work, we explore the use of synthetic data generated by a VC model to assist the TTS model in cross-speaker style transfer tasks. Additionally, we employ pre-training of the style encoder using timbre perturbation and prototypical angular loss to mitigate speaker leakage. Our results show that using VC synthetic data can improve the naturalness and speaker similarity of TTS in cross-speaker scenarios. Furthermore, we extend this approach to a cross-language scenario, enhancing accent transfer.

\end{abstract}

\section{Introduction}

Generating expressive speech in text-to-speech (TTS) models usually relies on recording and using data with specific expressive styles. However, recording for all desired speakers is not feasible in many cases. Recent studies aim to mitigate this problem through style transfer between speakers (cross-speaker style transfer), where the speaking style of a reference speaker is transferred to a target speaker with only neutral recordings.

In~\cite{skerry-ryan_towards_2018}, the Reference Encoder (RE) is proposed as a module that generates a style representation extracted from a reference audio and is incorporated into the TTS model, enabling prosody transfer from a reference speaker to a target speaker. When dealing with datasets labeled with styles, approaches such as \cite{kwon_effective_2019} and \cite{sorin_principal_2020} use these style representations to generate a latent space of representations, thereby controlling and transferring speech styles to neutral speakers in a cross-speaker approach. While the first one uses the centroids of each emotion to perform cross-speaker style transfer, the second employs Principal Component Analysis (PCA) to achieve the same.

These approaches, based on style representations generated by a module coupled with the TTS model, known as style encoder, rely on generating this latent representation from a reference mel-spectrogram, usually the TTS target mel-spectrogram during training. These modules aim to create a bottleneck capable of capturing prosodic information from the reference while ignoring content or speaker-specific information. In practice, this approach tends to leak speaker information into these representations (speaker leakage), causing the synthetic speech's timbre to match that of the conditioning representation during inference~\cite{skerry-ryan_towards_2018, valles-perez_improving_2021}. Moreover, degraded performance is observed when the trained TTS tries to transfer a specific style to a new speaker~\cite{an_disentangling_2022}.

Recent approaches aim to propose style encoders capable of generating disentangled representations, thereby avoiding speaker leakage. In \cite{shang_incorporating_2021} and \cite{zaidi_2022_daft} , the use of a gradient reversal layer (GRL)~\cite{ganin_unsupervised_2015} is proposed to remove speaker information from style encoder. However, these methods still suffer from source speaker leakage and low perceived naturalness in cross-speaker scenarios~\cite{sigurgeirsson_prosody_2023}.

In Nansy~\cite{choi_neural_2021}, formant shifting is used to remove speaker information from one of the model's modules. By perturbing the speaker information (timbre perturbation) of this module, it was able to model aspects unrelated to the speaker's timbre. This approach was utilized by \cite{lei_cross-speaker_2022} to disentangle speaker information and emotion, thereby transferring the style of the source speaker while maintaining the target speaker's timbre. In \cite{zhu_metts_2024}, timbre perturbation was also employed, successfully transferring not only emotions but also accents, proposing a model capable of performing cross-speaker style and accent transfer.

In scenarios where a large amount of expressive data is not easily available, it may affect the model's ability to perform style transfer. The use of synthetic data to incorporate cross-speaker style transfer in low-resource scenarios was proposed in \cite{huybrechts_lowresource_2021}, where a voice conversion (VC) model trained on neutral target speakers is used to convert source expressive speech (newscaster and conversational). The TTS is then fine-tuned with the expressive synthetic target speaker data. \cite{sam_ribeiro_cross-speaker_2022} also applies a similar methodology to transfer a conversational style to neutral speakers. In \cite{yoon_enhancingmultilingual_2024}, voice conversion is used to generate synthetic cross-language data, as VC can effectively retain content information, thereby training a multilingual TTS system. Although several approaches use synthetic data in TTS training, the quality of the VC model is extremely important, especially in low-resource and highly expressive data~\cite{terashima_cross-speaker_2022}.

In this work, we explore the use of synthetic data generated by an F0-conditioned VITS-based~\cite{kim_conditional_2021} voice conversion model to create expressive data for neutral speakers' voices. We conducted experiments using synthetic data combined with original data to achieve cross-speaker style transfer in a TTS model based on FastPitch~\cite{lancucki_fastpitch_2021}, augmented with a RE-based style encoder. To obtain meaningful style representations from the style encoder and avoid speaker leakage, we pre-trained it separately using both timbre perturbation and metric learning. Our observations indicate that synthetic data can enhance both naturalness and speaker similarity in cross-speaker TTS models. Additionally, by conditioning on original style representations, it is possible to maintain naturalness improvement while performing style transfer, even for styles that the VC model could not transfer well. Furthermore, we demonstrate that this approach is effective for accent transfer in cross-language tasks. Audio samples are provided in demo page\footnote{Audio samples available at: \url{http://bit.ly/3VCzFdd}}.

\section{Method}

\begin{figure}[t]
  \centering
  \includegraphics[width=\linewidth]{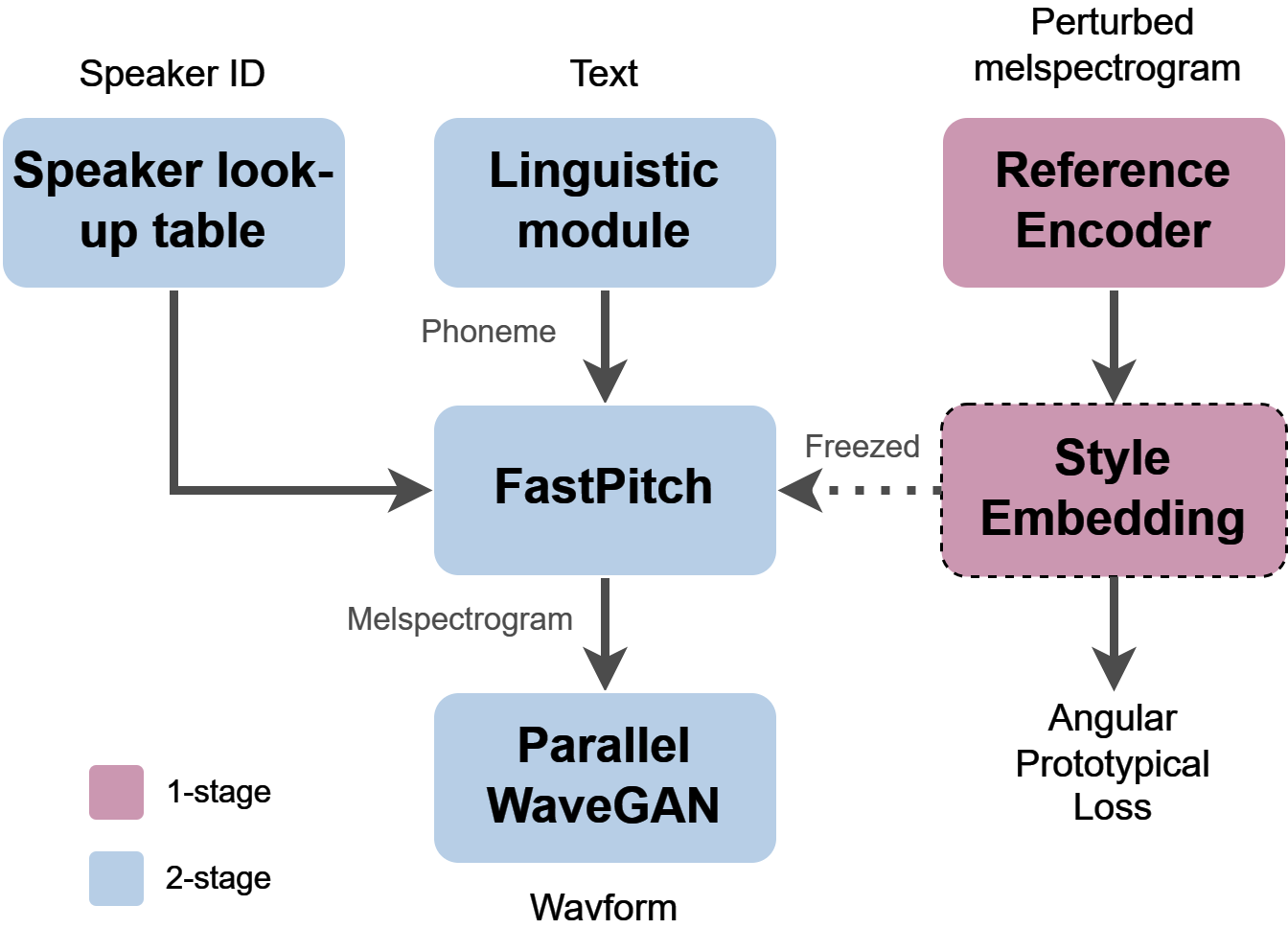}
  \caption{Two stages TTS training pipeline. Style encoder is pre-trained in stage 1 and remains frozen during TTS training at stage 2. The speaker look-up table encodes speaker information in FastPitch for multi-speaker extension.}
  \label{fig:trainingpipeline}
\end{figure}

We explore the use of synthetic data to aid in cross-speaker style transfer in a TTS model with a style encoder. Initially, we train a voice conversion model on neutral data and convert the expressive speech of the source speaker to all target speakers. Next, we use a two-stage approach for training the TTS model (Figure~\ref{fig:trainingpipeline}). In the first stage, we pre-train the style encoder, and in the second stage, we train the TTS using the pre-trained, frozen style encoder. In all stages, the original data is used with or without the addition of synthetic data. To explore the use of synthetic data, we conduct three experiments: 

\begin{itemize}
    \item \textbf{Synth none}: Only ground truth data is used;
    \item \textbf{Synth TTS}: Augmenting with synthetic data only in TTS training (stage 2);
    \item \textbf{Synth both}: Augmenting with synthetic data in both style encoder and TTS (stages 1 and 2).
\end{itemize}

\subsection{Synthetic data}

To generate synthetic data, we employed the open-source SO-VITS-SVC\footnote{SO-VITS-SVC available at \url{github.com/PlayVoice/whisper-vits-svc}} F0-conditioned voice conversion model. This model integrates four distinct audio encoders, each extracting different meaningful representations. A pre-trained timbre encoder extracts speaker embeddings, a Whisper~\cite{radford_2023_whisper} encoder extracts content information, a soft HuBERT~\cite{van2022comparison} model extracts prosody representation, and a CREPE~\cite{kim_2018_crepe} model extracts the F0. These representations are processed by a flow-based decoder and an adversarial training process based on VITS~\cite{kim_conditional_2021}. The model is also trained with a speaker classifier using a gradient reversal layer to achieve speaker disentanglement. We used the publicly available pre-trained checkpoint to fine-tune on our datasets. To ensure consistent conversions in cross-speaker scenarios, we also extracted the F0 statistics for each speaker in the dataset, and at inference time, we performed semitonal correction on the source expressive speech F0 to match the target speaker's F0.

\subsection{Stage 1: Style Encoder}

For the style encoder, we used the Reference Encoder (RE)~\cite{skerry-ryan_towards_2018} with several modifications to prevent information leakage. First, it was pre-trained separately to avoid modeling timbre information during TTS training. Additionally, to generate representations with distinct clusters for each style, we trained the Reference Encoder using the Prototypical Angular Loss~\cite{chung_defence_2020}.

We also perturbed the input mel-spectrogram of the Reference Encoder using two approaches. The first approach randomly selects slices of frames from the mel-spectrogram, ensuring that different segments of speech with the same style are used in each iteration. The second approach applies the timbre perturbation proposed by \cite{choi_neural_2021}, which involves randomly applying formant shifting to the input audio and then passing the perturbed mel-spectrogram to the Reference Encoder.

\subsection{Stage 2: TTS}

For the TTS model, we based our approach on the FastPitch~\cite{lancucki_fastpitch_2021} architecture, which is a non-autoregressive model conditioned on F0. We also included speech energy modeling and normalized both energy and F0 to the statistics of each speaker in the dataset to reduce the amount of speaker-specific information carried by these features. During training, FastPitch is conditioned on the representations generated by the pre-trained Reference Encoder with its layers frozen. During inference, the ground truth centroid representation of each style is used to perform cross-speaker style transfer.

To achieve faster convergence, we trained all TTS models from a pre-trained FastPitch model trained on six neutral Brazilian Portuguese speakers from an 30h internal dataset. Finally, we used the Parallel WaveGAN~\cite{yamamoto_parallel_2020} vocoder to convert the mel-spectrogram into waveform.

\section{Experiments}

We performed the experiments on an internal Brazilian Portuguese dataset. The dataset consists of three speakers, coded as PTBR 1 (female), PTBR 2 (male), and PTBR 3 (female), as shown in Table~\ref{tab:stylecpqd_dataset}. Speaker PTBR 1 has three highly expressive styles of speech in addition to neutral: lively, welcoming, and harsh. Speakers PTBR 2 and 3 only have neutral recordings.

\begin{table}[b!]
\centering
\caption{Training dataset detailed overview.}
\label{tab:stylecpqd_dataset}
    \resizebox{0.5\textwidth}{!}{%
    \begin{tabular}{c|c|c|c|c}
    \textbf{Speaker ID} & \textbf{Style (pt-br)} & \textbf{Style (en)} & \textbf{\#Files} & \textbf{\#Hours} \\ \hline
        \multirow{4}{4em}{PTBR~1} & Neutro & Neutral & 3574  & 3.97 \\ 
         & Animado & Lively & 1307  & 1.66 \\ 
         & Acolhedor & Welcoming & 1308  & 1.69 \\ 
         & Rispido & Harsh & 1256  & 1.60 \\ \hline
        PTBR~2 & Neutro & Neutral & 4725  & 4.62 \\ \hline
        PTBR~3 & Neutro & Neutral & 6759  & 7.29 \\ \hline \hline
    & & & 18929 & 20.83
    \end{tabular}%
            }
\end{table}

We used 90\% of the data for training, 10\% for validation, and manually selected 20 style-paired samples from PTBR 1 and 20 random samples for each remaining speaker for evaluation.

Additionally, to evaluate this approach for cross-language in a one-speaker-one-language scenario (only one speaker for each foreign language), we conducted experiments aimed at transferring native English and Spanish accents to Brazilian Portuguese speakers. We used the LJspeech dataset for English references and the Blizzard2021 dataset for Spanish references. We used 80\% of each dataset for training, 10\% for validation, and manually extracted 20 out-of-distribution sentences from each language for testing.

\subsection{Training setup}

All experiments were conducted on a single T4 GPU (16GB) using the same training and validation partitions. The voice conversion model was trained for 100k steps starting from a publicly available checkpoint with a batch size of 8 with 2 steps of gradient accumulation and AdamW with learning rate (LR) of 5e-5. The style encoder was trained from scratch for 60k steps, with spectrogram segments of 1.6s, a formant shifting factor of 1.4, a batch of 10 examples for each of the 4 styles in the dataset, resulting in a total batch size of 40 and RAdam with LR 1e-4. The TTS models were trained from a pre-trained neutral model for 250k steps with a batch size of 16 and Adam with LR 1e-4. The Parallel WaveGAN vocoder was trained in a 30h internal dataset (PTBR speakers included) for 600k steps from scratch using the recipe from Parallel WaveGAN repository\footnote{PW-GAN available at: \url{https://github.com/kan-bayashi/ParallelWaveGAN}}.

\subsection{Evaluation}

We evaluated the model through subjective assessments of naturalness, style intensity, and speaker similarity. A total of 21 native Brazilian participants conducted the tests. For naturalness, we used a Mean Opinion Score (MOS) evaluation ranging from 1 to 5, where 1 represented completely artificial audio and 5 represented completely natural audio. Each participant evaluated 5 sentences generated by each of the models. The sentences were generated for each voice and style in the dataset, totaling 12 stimuli per sentence for each model.

To assess the intensity of emotion, an example page with samples of each style was presented to describe each style in the dataset. The evaluation used a MOS scale where 1 was given to audio completely different from the target style, and higher scores represented increasing intensities of the audio exhibiting that style, with 5 being the highest intensity. For each expressive style in the dataset, 3 sentences were evaluated in each of the 3 voices by each model, totaling 9 stimuli per model for each style.

Finally, for speaker similarity, 4 sentences for each speaker were evaluated with stimuli generated by each of the explored models. On each evaluation page, a ground truth reference of the target voice was provided, and the participants rated how similar each stimulus was to the reference voice, with 1 meaning it did not resemble the reference at all and 5 meaning it was the same person speaking.

For the cross-language experiment, we assessed the results through objective metrics. We used UTMOS~\cite{saeki22c_interspeech} to estimate the naturalness of synthetic speech, character error rate (CER) for intelligibility evaluation using Whisper large, and a Resemblyzer\footnote{Resemblyzer available at: \url{https://github.com/resemble-ai/Resemblyzer}}-based speaker embedding cosine similarity (SECS) metric to assess speaker similarity, similar to used in \cite{kang_zet-speech_2023}.

\section{Results}

We report the overall naturalness MOS results in Table~\ref{tab:naturalness}. The results indicate that synthetic data generated by voice conversion (VC) exhibit higher naturalness than those from cross-speaker style transfer experiments. Consequently, both the Synth TTS and Synth both experiments, which incorporate synthetic data during training, show increased naturalness compared to Synth None, with Synth both demonstrating the highest improvement.

\begin{table}[h]
  \caption{Naturalness MOS with 95\% confidence intervals.}
  \label{tab:naturalness}
  \centering
  \begin{tabular}{cc}
    \toprule
    \multicolumn{1}{c}{\textbf{Model}} & 
    \multicolumn{1}{c}{\textbf{MOS}} \\ 
    \midrule
    GT & $4.11\pm0.13$ \\
    Synth none & $2.53\pm0.13$ \\
    Synth TTS & $2.79\pm0.13$ \\
    Synth both & \underline{$3.06\pm0.14$}  \\
    VC & $\mathbf{3.78\pm0.20}$  \\
    \bottomrule
  \end{tabular}
\end{table}

\begin{table*}[ht!]
  \caption{Style Intensity (SI-MOS) and Naturalness (N-MOS) Mean Opinion Scores with 95\% confidence intervals.}
  \label{tab:style}
  \centering
  \begin{tabular}{ccccccc}
    \toprule
    \multicolumn{1}{c}{\textbf{}} &  \multicolumn{2}{c}{\textbf{Lively}}&  \multicolumn{2}{c}{\textbf{Harsh}}&  \multicolumn{2}{c}{\textbf{Wellcoming}} \\ 
    \textbf{Model} & SI-MOS & N-MOS& SI-MOS & N-MOS& SI-MOS & N-MOS \\
    \midrule
    GT & $4.49\pm0.14$ & $4.28\pm0.38$ & $4.52\pm0.14$ & $4.85\pm0.21$ & $4.01\pm0.26$ & $4.28\pm0.29$  \\
    Synth none & \underline{$2.67\pm0.26$} & $1.74\pm0.24$ & $\mathbf{2.71\pm0.14}$ & $2.20\pm0.22$ & $2.63\pm0.17$ & $2.33\pm0.27$ \\
    Synth TTS & $2.65\pm0.15$ & $2.55\pm0.16$ & $2.17\pm0.15$ & $2.11\pm0.19$ & $2.99\pm0.14$ & $2.55\pm0.27$  \\
    Synth both & $2.24\pm0.14$ & \underline{$2.90\pm0.31$} & \underline{$2.25\pm0.15$} & \underline{$2.28\pm0.20$} & \underline{$3.06\pm0.17$} & \underline{$3.44\pm0.28$} \\
    VC & $\mathbf{3.04\pm0.16}$ & $\mathbf{3.66\pm0.26}$ & $1.93\pm0.29$ & $\mathbf{3.66\pm0.38}$ & $\mathbf{3.61\pm0.19}$ & $\mathbf{3.95\pm0.54}$   \\
    \bottomrule
  \end{tabular}
\end{table*}

However, we note that for style intensity, the Synth none configuration performed better in two out of three expressive styles in the dataset, despite having lower naturalness in each case (Table~\ref{tab:style}).

Although VC can achieve higher overall naturalness for all styles, it drops in terms of style intensity for specific styles. Specifically, in the harsh style, Synth None exhibits higher style intensity than VC, even though it performs lower in terms of naturalness in this style. Synth both consistently maintains a mid-level of higher style intensity and higher naturalness. In cases where VC's style intensity is higher (e.g., welcoming), Synth both achieves higher style intensity than the other configurations. Moreover, even in the harsh style, where VC performs poorly, Synth both can still retain some style intensity from the original data.

Regarding speaker similarity (Table~\ref{tab:speaker}), we observe that Synth TTS mostly preserves the similarity of timbre in cross-speaker scenarios, with Synth both following closely. However, Synth none performs worse in this aspect.

\begin{table}[ht!]
  \caption{Speaker Similarity MOS 95\% confidence intervals.}
  \label{tab:speaker}
  \centering
  \begin{tabular}{cc}
    \toprule
    \multicolumn{1}{c}{\textbf{Model}} & 
    \multicolumn{1}{c}{\textbf{Similarity}} \\ 
    \midrule
    Synth none & $2.44\pm0.20$ \\
    Synth TTS & $\mathbf{2.93\pm0.20}$ \\
    Synth both & \underline{$2.72\pm0.19$}  \\
    \bottomrule
  \end{tabular}
\end{table}

\begin{figure*}[ht]
  \centering
  \includegraphics[width=0.8\linewidth]{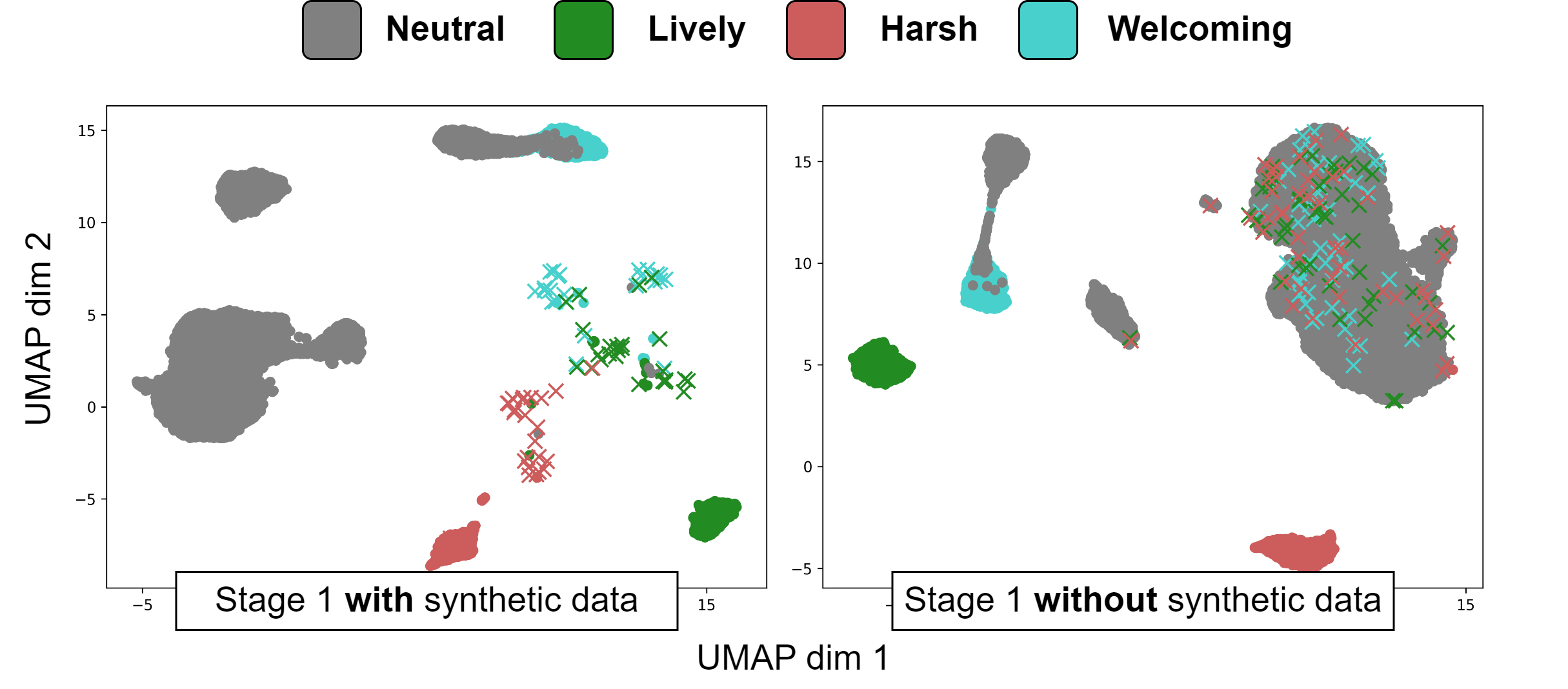}
  \caption{Style representations projected using UMAP~\cite{McInnes2018} when training with and without synthetic data in Stage 1 (``x'' markers are synthetic expressive data).}
  \label{fig:stylespace}
\end{figure*}

Although using only original expressive data appears to yield higher style intensity perception in synthetic speech, it results in lower naturalness and speaker similarity. Employing synthetic data generated by a Voice Conversion (VC) model improved both naturalness and speaker similarity. However, the effectiveness of style transfer in TTS for highly expressive styles is sensitive to the quality of the VC for each style. Also, training stage 1 with both synthetic and ground truth data generates a more meaningful representations as shown in Figure~\ref{fig:stylespace}.

When synthetic data is not used in style encoder training, the expressive synthetic data share the same representation as the original neutral data. While this does not affect the overall cross-speaker style transfer in experiments, since the ground truth centroids are used for inference, it prevents the TTS model from leveraging the style information of the synthetic samples.

Analyzing the effect of using synthetic data in a more content-dependent cross-speaker transfer, we performed the same experiments to transfer accents for our three PTBR speakers to two languages: English and Spanish. We report the objective metrics in Table~\ref{tab:languages}.

\begin{table}[ht]
  \caption{Cross-language objective metrics.}
  \label{tab:languages}
  \centering
  \begin{tabular}{cccc}
    \toprule
    \multicolumn{1}{c}{\textbf{Model}} & 
    \multicolumn{1}{c}{\textbf{UTMOS $\uparrow$}} & 
    \multicolumn{1}{c}{\textbf{CER (\%) $\downarrow$}} & 
    \multicolumn{1}{c}{\textbf{SECS} $\uparrow$}\\ 
    \midrule
    GT & 4.03 & - &  - \\
    Synth none & 2.93 & 12.73 & 0.56\\
    Synth TTS & 3.04 & 10.42 & 0.59\\
    Synth both & \textbf{3.32} & \textbf{10.19} & \textbf{0.79} \\
    \bottomrule
  \end{tabular}
\end{table}

We observed that in this case, Synth both performed better than the other methods. Specifically, not using synthetic data appears to suffer from both content and speaker leakage. Adding synthetic data only in TTS seems to improve overall naturalness and intelligibility, but it still suffers from speaker leakage. Finally, by combining synthetic speech in both style encoder pre-training and TTS, it is possible to disentangle speaker information from accent representations, allowing the model to transfer the accent in a cross-speaker scenario, even when each language is dependent on a specific speaker.

\section{Conclusion and future work}

Cross-speaker style transfer in low resource expressive data scenarios is still hard due to quality degradation or information leakage. Results shows that synthetic data can be used to improve both naturalness and speaker similarity in style transfer scenario. Although style intensity is very sensitive to the quality of the style transfer capacity of voice conversion model, using it as partial representations together with ground truth style data can balance style and naturalness of synthetic speech. This approach seems effective for cross-language accent-transfer even in scenario where we have only one speaker for each foreign language, improving naturalness, intelligibility and speaker similarity. In future work, we intend to explore the usage of cross-language expressive datasets to improve languages with low resource expressive data through cross-language cross-speaker synthetic augmentation.

\section{Acknowledgements}
This study is partially funded by the Coordenação de Aperfeiçoamento de Pessoal de Nivel Superior – Brasil (CAPES) – Finance Code 001, and it is supported by the BI0S - Brazilian Institute of Data Science, grant \#2020/09838-0, São Paulo Research Foundation (FAPESP). Paula D. P. Costa, Lucas H. Ueda, and Leonardo B. de M. M. Marques are also affiliated with the Dept. of Computer Engineering and Automation (DCA), Faculdade de Engenharia Elétrica e de Computação and are part of the Artificial Intelligence Lab., Recod.ai, Institute of Computing, UNICAMP.

\bibliographystyle{IEEEtran}
\bibliography{template}

\begin{thebibliography}{10}
\providecommand{\url}[1]{#1}
\csname url@samestyle\endcsname
\providecommand{\newblock}{\relax}
\providecommand{\bibinfo}[2]{#2}
\providecommand{\BIBentrySTDinterwordspacing}{\spaceskip=0pt\relax}
\providecommand{\BIBentryALTinterwordstretchfactor}{4}
\providecommand{\BIBentryALTinterwordspacing}{\spaceskip=\fontdimen2\font plus
\BIBentryALTinterwordstretchfactor\fontdimen3\font minus \fontdimen4\font\relax}
\providecommand{\BIBforeignlanguage}[2]{{%
\expandafter\ifx\csname l@#1\endcsname\relax
\typeout{** WARNING: IEEEtran.bst: No hyphenation pattern has been}%
\typeout{** loaded for the language `#1'. Using the pattern for}%
\typeout{** the default language instead.}%
\else
\language=\csname l@#1\endcsname
\fi
#2}}
\providecommand{\BIBdecl}{\relax}
\BIBdecl

\bibitem{skerry-ryan_towards_2018}
\BIBentryALTinterwordspacing
R.~J. Skerry-Ryan, E.~Battenberg, Y.~Xiao, Y.~Wang, D.~Stanton, J.~Shor, R.~Weiss, R.~Clark, and R.~A. Saurous, ``\BIBforeignlanguage{en}{Towards {End}-to-{End} {Prosody} {Transfer} for {Expressive} {Speech} {Synthesis} with {Tacotron}},'' in \emph{\BIBforeignlanguage{en}{Proceedings of the 35th {International} {Conference} on {Machine} {Learning}}}.\hskip 1em plus 0.5em minus 0.4em\relax PMLR, Jul. 2018, pp. 4693--4702, iSSN: 2640-3498. [Online]. Available: \url{https://proceedings.mlr.press/v80/skerry-ryan18a.html}
\BIBentrySTDinterwordspacing

\bibitem{kwon_effective_2019}
\BIBentryALTinterwordspacing
O.~Kwon, I.~Jang, C.~Ahn, and H.-G. Kang, ``An {Effective} {Style} {Token} {Weight} {Control} {Technique} for {End}-to-{End} {Emotional} {Speech} {Synthesis},'' \emph{IEEE Signal Processing Letters}, vol.~26, no.~9, pp. 1383--1387, Sep. 2019, conference Name: IEEE Signal Processing Letters. [Online]. Available: \url{https://ieeexplore.ieee.org/document/8778667}
\BIBentrySTDinterwordspacing

\bibitem{sorin_principal_2020}
\BIBentryALTinterwordspacing
A.~Sorin, S.~Shechtman, and R.~Hoory, ``Principal {Style} {Components}: {Expressive} {Style} {Control} and {Cross}-{Speaker} {Transfer} in {Neural} {TTS},'' 2020, pp. 3411--3415. [Online]. Available: \url{https://www.isca-archive.org/interspeech_2020/sorin20_interspeech.html}
\BIBentrySTDinterwordspacing

\bibitem{valles-perez_improving_2021}
\BIBentryALTinterwordspacing
I.~Vallés-Pérez, J.~Roth, G.~Beringer, R.~Barra-Chicote, and J.~Droppo, ``\BIBforeignlanguage{en}{Improving multi-speaker {TTS} prosody variance with a residual encoder and normalizing flows},'' Jun. 2021, arXiv:2106.05762 [cs, eess]. [Online]. Available: \url{http://arxiv.org/abs/2106.05762}
\BIBentrySTDinterwordspacing

\bibitem{an_disentangling_2022}
\BIBentryALTinterwordspacing
X.~An, F.~K. Soong, and L.~Xie, ``Disentangling {Style} and {Speaker} {Attributes} for {TTS} {Style} {Transfer},'' \emph{IEEE/ACM Transactions on Audio, Speech, and Language Processing}, vol.~30, pp. 646--658, 2022, conference Name: IEEE/ACM Transactions on Audio, Speech, and Language Processing. [Online]. Available: \url{https://ieeexplore.ieee.org/document/9693198/}
\BIBentrySTDinterwordspacing

\bibitem{shang_incorporating_2021}
Z.~Shang, Z.~Huang, H.~Zhang, P.~Zhang, and Y.~Yan, \emph{Incorporating {Cross}-{Speaker} {Style} {Transfer} for {Multi}-{Language} {Text}-to-{Speech}}, Aug. 2021, pages: 1623.

\bibitem{zaidi_2022_daft}
J.~Zaïdi, H.~Seuté, B.~{van Niekerk}, and M.-A. Carbonneau, ``{Daft-Exprt: Cross-Speaker Prosody Transfer on Any Text for Expressive Speech Synthesis},'' in \emph{Proc. Interspeech 2022}, 2022, pp. 4591--4595.

\bibitem{ganin_unsupervised_2015}
Y.~Ganin and V.~Lempitsky, ``\BIBforeignlanguage{en}{Unsupervised {Domain} {Adaptation} by {Backpropagation}},'' 2015.

\bibitem{sigurgeirsson_prosody_2023}
\BIBentryALTinterwordspacing
A.~T. Sigurgeirsson and S.~King, ``\BIBforeignlanguage{en}{Do {Prosody} {Transfer} {Models} {Transfer} {Prosody}?}'' Mar. 2023, arXiv:2303.04289 [cs, eess]. [Online]. Available: \url{http://arxiv.org/abs/2303.04289}
\BIBentrySTDinterwordspacing

\bibitem{choi_neural_2021}
\BIBentryALTinterwordspacing
H.-S. Choi, J.~Lee, W.~Kim, J.~Lee, H.~Heo, and K.~Lee, ``Neural {Analysis} and {Synthesis}: {Reconstructing} {Speech} from {Self}-{Supervised} {Representations},'' in \emph{Advances in {Neural} {Information} {Processing} {Systems}}, vol.~34.\hskip 1em plus 0.5em minus 0.4em\relax Curran Associates, Inc., 2021, pp. 16\,251--16\,265. [Online]. Available: \url{https://proceedings.neurips.cc/paper_files/paper/2021/hash/87682805257e619d49b8e0dfdc14affa-Abstract.html}
\BIBentrySTDinterwordspacing

\bibitem{lei_cross-speaker_2022}
\BIBentryALTinterwordspacing
Y.~Lei, S.~Yang, X.~Zhu, L.~Xie, and D.~Su, ``Cross-{Speaker} {Emotion} {Transfer} {Through} {Information} {Perturbation} in {Emotional} {Speech} {Synthesis},'' \emph{IEEE Signal Processing Letters}, vol.~29, pp. 1948--1952, 2022, conference Name: IEEE Signal Processing Letters. [Online]. Available: \url{https://ieeexplore.ieee.org/document/9874835}
\BIBentrySTDinterwordspacing

\bibitem{zhu_metts_2024}
\BIBentryALTinterwordspacing
X.~Zhu, Y.~Lei, T.~Li, Y.~Zhang, H.~Zhou, H.~Lu, and L.~Xie, ``{METTS}: {Multilingual} {Emotional} {Text}-to-{Speech} by {Cross}-{Speaker} and {Cross}-{Lingual} {Emotion} {Transfer},'' \emph{IEEE/ACM Transactions on Audio, Speech and Language Processing}, vol.~32, pp. 1506--1518, Feb. 2024. [Online]. Available: \url{https://doi.org/10.1109/TASLP.2024.3363444}
\BIBentrySTDinterwordspacing

\bibitem{huybrechts_lowresource_2021}
G.~Huybrechts, T.~Merritt, G.~Comini, B.~Perz, R.~Shah, and J.~Lorenzo-Trueba, ``Low-resource expressive text-to-speech using data augmentation,'' in \emph{ICASSP 2021 - 2021 IEEE International Conference on Acoustics, Speech and Signal Processing (ICASSP)}, 2021, pp. 6593--6597.

\bibitem{sam_ribeiro_cross-speaker_2022}
\BIBentryALTinterwordspacing
M.~Sam~Ribeiro, J.~Roth, G.~Comini, G.~Huybrechts, A.~Gabryś, and J.~Lorenzo-Trueba, ``Cross-{Speaker} {Style} {Transfer} for {Text}-to-{Speech} {Using} {Data} {Augmentation},'' in \emph{{ICASSP} 2022 - 2022 {IEEE} {International} {Conference} on {Acoustics}, {Speech} and {Signal} {Processing} ({ICASSP})}, May 2022, pp. 6797--6801, iSSN: 2379-190X. [Online]. Available: \url{https://ieeexplore.ieee.org/document/9746179}
\BIBentrySTDinterwordspacing

\bibitem{yoon_enhancingmultilingual_2024}
H.-W. Yoon, J.-S. Kim, R.~Yamamoto, R.~Terashima, C.-H. Song, J.-M. Kim, and E.~Song, ``Enhancing multilingual tts with voice conversion based data augmentation and posterior embedding,'' in \emph{ICASSP 2024 - 2024 IEEE International Conference on Acoustics, Speech and Signal Processing (ICASSP)}, 2024, pp. 12\,186--12\,190.

\bibitem{terashima_cross-speaker_2022}
\BIBentryALTinterwordspacing
R.~Terashima, R.~Yamamoto, E.~Song, Y.~Shirahata, H.-W. Yoon, J.-M. Kim, and K.~Tachibana, ``\BIBforeignlanguage{en}{Cross-{Speaker} {Emotion} {Transfer} for {Low}-{Resource} {Text}-to-{Speech} {Using} {Non}-{Parallel} {Voice} {Conversion} with {Pitch}-{Shift} {Data} {Augmentation}},'' in \emph{\BIBforeignlanguage{en}{Interspeech 2022}}.\hskip 1em plus 0.5em minus 0.4em\relax ISCA, Sep. 2022, pp. 3018--3022. [Online]. Available: \url{https://www.isca-archive.org/interspeech_2022/terashima22_interspeech.html}
\BIBentrySTDinterwordspacing

\bibitem{kim_conditional_2021}
\BIBentryALTinterwordspacing
J.~Kim, J.~Kong, and J.~Son, ``\BIBforeignlanguage{en}{Conditional {Variational} {Autoencoder} with {Adversarial} {Learning} for {End}-to-{End} {Text}-to-{Speech}},'' in \emph{\BIBforeignlanguage{en}{Proceedings of the 38th {International} {Conference} on {Machine} {Learning}}}.\hskip 1em plus 0.5em minus 0.4em\relax PMLR, Jul. 2021, pp. 5530--5540, iSSN: 2640-3498. [Online]. Available: \url{https://proceedings.mlr.press/v139/kim21f.html}
\BIBentrySTDinterwordspacing

\bibitem{lancucki_fastpitch_2021}
\BIBentryALTinterwordspacing
A.~Łańcucki, ``Fastpitch: {Parallel} {Text}-to-{Speech} with {Pitch} {Prediction},'' in \emph{{ICASSP} 2021 - 2021 {IEEE} {International} {Conference} on {Acoustics}, {Speech} and {Signal} {Processing} ({ICASSP})}, Jun. 2021, pp. 6588--6592, iSSN: 2379-190X. [Online]. Available: \url{https://ieeexplore.ieee.org/document/9413889}
\BIBentrySTDinterwordspacing

\bibitem{radford_2023_whisper}
\BIBentryALTinterwordspacing
A.~Radford, J.~W. Kim, T.~Xu, G.~Brockman, C.~Mcleavey, and I.~Sutskever, ``Robust speech recognition via large-scale weak supervision,'' in \emph{Proceedings of the 40th International Conference on Machine Learning}, ser. Proceedings of Machine Learning Research, A.~Krause, E.~Brunskill, K.~Cho, B.~Engelhardt, S.~Sabato, and J.~Scarlett, Eds., vol. 202.\hskip 1em plus 0.5em minus 0.4em\relax PMLR, 23--29 Jul 2023, pp. 28\,492--28\,518. [Online]. Available: \url{https://proceedings.mlr.press/v202/radford23a.html}
\BIBentrySTDinterwordspacing

\bibitem{van2022comparison}
B.~van Niekerk, M.-A. Carbonneau, J.~Za{\"\i}di, M.~Baas, H.~Seut{\'e}, and H.~Kamper, ``A comparison of discrete and soft speech units for improved voice conversion,'' in \emph{ICASSP 2022-2022 IEEE International Conference on Acoustics, Speech and Signal Processing (ICASSP)}.\hskip 1em plus 0.5em minus 0.4em\relax IEEE, 2022, pp. 6562--6566.

\bibitem{kim_2018_crepe}
J.~W. Kim, J.~Salamon, P.~Li, and J.~P. Bello, ``Crepe: A convolutional representation for pitch estimation,'' in \emph{2018 IEEE International Conference on Acoustics, Speech and Signal Processing (ICASSP)}, 2018, pp. 161--165.

\bibitem{chung_defence_2020}
\BIBentryALTinterwordspacing
J.~S. Chung, J.~Huh, S.~Mun, M.~Lee, H.-S. Heo, S.~Choe, C.~Ham, S.~Jung, B.-J. Lee, and I.~Han, ``In {Defence} of {Metric} {Learning} for {Speaker} {Recognition},'' 2020, pp. 2977--2981. [Online]. Available: \url{https://www.isca-archive.org/interspeech_2020/chung20b_interspeech.html}
\BIBentrySTDinterwordspacing

\bibitem{yamamoto_parallel_2020}
\BIBentryALTinterwordspacing
R.~Yamamoto, E.~Song, and J.-M. Kim, ``Parallel {Wavegan}: {A} {Fast} {Waveform} {Generation} {Model} {Based} on {Generative} {Adversarial} {Networks} with {Multi}-{Resolution} {Spectrogram},'' in \emph{{ICASSP} 2020 - 2020 {IEEE} {International} {Conference} on {Acoustics}, {Speech} and {Signal} {Processing} ({ICASSP})}, May 2020, pp. 6199--6203, iSSN: 2379-190X. [Online]. Available: \url{https://ieeexplore.ieee.org/document/9053795}
\BIBentrySTDinterwordspacing

\bibitem{saeki22c_interspeech}
T.~Saeki, D.~Xin, W.~Nakata, T.~Koriyama, S.~Takamichi, and H.~Saruwatari, ``{UTMOS: UTokyo-SaruLab System for VoiceMOS Challenge 2022},'' in \emph{Proc. Interspeech 2022}, 2022, pp. 4521--4525.

\bibitem{kang_zet-speech_2023}
\BIBentryALTinterwordspacing
M.~Kang, W.~Han, S.~J. Hwang, and E.~Yang, ``\BIBforeignlanguage{en}{{ZET}-{Speech}: {Zero}-shot adaptive {Emotion}-controllable {Text}-to-{Speech} {Synthesis} with {Diffusion} and {Style}-based {Models}},'' in \emph{\BIBforeignlanguage{en}{{INTERSPEECH} 2023}}.\hskip 1em plus 0.5em minus 0.4em\relax ISCA, Aug. 2023, pp. 4339--4343. [Online]. Available: \url{https://www.isca-archive.org/interspeech_2023/kang23_interspeech.html}
\BIBentrySTDinterwordspacing

\bibitem{McInnes2018}
\BIBentryALTinterwordspacing
L.~McInnes, J.~Healy, N.~Saul, and L.~Großberger, ``Umap: Uniform manifold approximation and projection,'' \emph{Journal of Open Source Software}, vol.~3, no.~29, p. 861, 2018. [Online]. Available: \url{https://doi.org/10.21105/joss.00861}
\BIBentrySTDinterwordspacing

\end{thebibliography}

\end{document}